\documentclass[doublecol]{epl2} 
\usepackage{epstopdf}
\usepackage{subfigure}

\title{Unjamming of Granular Packings as a Constraint Satisfaction Problem:
Evidence for a  Growing Static Length Scale in Frictionless Packings}
\shorttitle{Growing Static Length Scale in Frictionless Packings} 

\author{M. Mailman\inst{1} \and B. Chakraborty\inst{1}}
\shortauthor{M. Mailman \etal}

\institute{                    
  \inst{1} Martin Fisher School of Physics, Brandeis University, Waltham, Massachusetts 02454}

\abstract{
An outstanding question in the physics of soft jammed packings concerns
the nature of the correlations that arise near the unjamming transition.
In this work, we treat unjamming as a constraint satisfaction problem and demonstrate that a  static correlation function, which probes sensitivity to boundary conditions,
exhibits
a diverging correlation length as the packing is decompressed. This length scale is related to isostaticity, and has been connected earlier to the existence of soft modes in isostatic packings. The form of the correlation function is remarkably similar to one predicted in the mosaic theory of the glass transition, and we relate the nature of the growing correlations to the entropy of packings.}

\begin{document}

\maketitle

\section{Introduction} There is a class of systems in soft-condensed matter, which exhibit pronounced anomalies in their dynamics without any dramatic changes in their static properties.   The viscosity of supercooled liquids increases dramatically as the temperature is lowered resulting in the liquid falling out of equilibrium at a nominal "glass transition" temperature\cite{BiroliNPhys}.  Systems that interact via purely repulsive, short-range forces, such as  granular materials can jam or unjam in response to external driving.   The dynamics in these systems become increasingly heterogeneous as the non-flowing state is approached, and there are remarkable similarities between the dynamical correlation 
functions of these two systems\cite{DurianDynLength,DauchotDynHet}, which are strikingly different in any standard measure.  Granular materials are athermal, and have frictional forces whereas supercooled liquids are in thermal equilibrium until they fall out of equilibrium at the glass transition temperature.   The one feature that connects these systems is the existence of many amorphous  metastable states\cite{BouchaudIdeas}.   

In mean-field theories of the glass transition such as the random-first-order theory,  a non-vanishing complexity or configurational entropy does lead to the definition of a static, point-to-set  (PTS) correlation  length that probes the growing influence of the boundary\cite{BiroliAdamGibbs,BiroliReview}.  The divergence of a PTS correlation length would indicate a true thermodynamic phase transition, where by definition boundary conditions affect the interior completely.  A question that naturally arises is whether such a PTS measure can be used to detect phase transitions in athermal systems, and in particular whether the jamming/unjamming transition in granular systems can be characterized by a similar correlation length.   In addition to its application to the glass transition problem, the PTS correlation function has been developed and widely used to  monitor the structure of the solution space of constraint satisfaction problems involving discrete or boolean variables such as those found in coloring problems and satisfiability problems,  for 
instance, XORSAT\cite{CSPKurchanKrzakala}.

Jamming and unjamming  can both be viewed as {}``constraint satisfaction''  problems.   If grains are idealized as being infinitely rigid, then the constraints for jamming are zero overlap between grains, and the degrees of freedom are the positions of the grains.   There has been considerable recent work analyzing the solution space of hard sphere packings as 
one approaches jamming, and there is evidence for a diverging length scale associated with this entropy-vanishing transition\cite{CSPKurchanKrzakala}. 
In systems interacting via soft repulsive potentials cohesion arises only as a consequence of compression, and a solid can unjam (fail mechanically) with all elastic moduli vanishing as the assembly of grains is decompressed\cite{JammingBible}.   For this type of unjamming the constraints are the equations of mechanical equilibrium, and  the degrees of freedom are the positions of the grains, and the contact forces.   Experiments in granular systems have shown that  unjamming through decompression is accompanied by large fluctuations 
of stress\cite{BehringerHowellCouette,BehringerTrush}.
It has been argued that for {\it frictionless} grains, the jammed configurations at this unjamming point, referred to as Point J\cite{Jamming98} are {\it critical}  with diverging correlation lengths, and soft modes\cite{CoreyCompressionRoutine,MatthieuLengthScale}.  So far, however, it has not been possible to measure a diverging correlation length from measurements of any static correlation functions\cite{JammingBible}.   The existence of a diverging length scale has been inferred from characteristics of the soft modes\cite{JammingBible}.   In this work, we demonstrate that the unjamming transition in two-dimensional assemblies of frictionless disks is characterized by a diverging  length in a PTS correlation function, and that this correlation function is intimately related to the entropy of jammed, mechanically stable states.  

\section{Model} In contrast to the constraint satisfaction problems that have been analyzed using the PTS correlation function, jamming/unjamming poses new challenges.  The degrees of freedom involved in unjamming are contact forces and positions of grains, which are continuous degrees of freedom.  Given the complexity of the problem, we study a minimal model of static granular packings known as the force-network-ensemble (FNE) \cite{FNE}, which  explicitly incorporates local force balance.  For hard but not infinitely rigid grains, small deformations can lead to large changes in the force magnitudes.  The length scales governing packing geometries and contact forces are, therefore, well separated \cite{FNE}.  In this limit, the jammed system can be modeled by the FNE, which is a flat distribution of all forces that satisfy the constraints of mechanical equilibrium on a given geometry\cite{FNE}.  The unjamming problem then reduces to a constraint satisfaction problem on a particular random network, where the degrees of freedom are the force magnitudes.  The FNE is a minimal model, which has been shown to capture the robust features of  contact force distributions, and mechanical properties of static granular packings.  Since we are interested in exploring correlations that are strong indicators of unjamming, and not sensitive to specific force laws, we study the PTS correlation function in FNE.  We measure PTS correlations as a function pressure (allowing the set of quenched geometries to change with changing pressure) and find that a static length scale does diverge as the system unjams. The origins of this length scale can be understood from a simple bulk-surface argument. In addition, we explore the structure of the solution space of the constraint satisfaction problem, and discuss its implications for the entropy of granular packings.  Parenthetically, the concept of the force-network ensemble is on firmer footing for frictional rather than frictionless grains since the indeterminacy exists even for infinitely rigid grains\cite{VanHeckeTigheReview}, and some of our conclusions could carry over to frictional systems.

To study unjamming, we need to sample a large set of amorphous geometries, and construct the corresponding FNEs. We use the O'Hern protocol\cite{CoreyCompressionRoutine} to generate an unbiased sampling of just-touching disk packings.  Packings at arbitrary overcompressions are then generated by increasing the packing fraction in a controlled manner. In this work, we have only studied packings obtained from a harmonic interaction.  The geometries obtained via this protocol are then used to construct the FNE.  For a given geometry, PTS is well defined as the overlap between force networks, and can then be averaged over various realizations of the quenched geometry. 

In a $d-$dimensional granular assembly of $M$ grains and 
$z$ contacts there are $\frac{z}{2}$
contact force variables, which are constrained by $dM$ mechanical
equilibrium (ME) equations. A minimum condition for finding valid
(i.e. satisfying ME) force networks is that $\frac{z}{2}\geq d M$.  
For overcompressed packings that satisfy this condition, FNE assumes a flat measure for the set of force
networks $\vec{f}$, which satisfy the matrix equation
\begin{equation}
A\vec{f}=\vec{b}\label{eq: matrix_eq}
\label{eq:matrix_eq}
\end{equation}
It is the nature of this solution space that will determine the structure of the PTS correlation function.  Throughout this work, we will refer to the vector $\vec{f}$ as a {}``force network," so as to avoid confusion with {}``force vector," which one typically associates with a particular force acting on a body.  The force network is a list of contact force magnitudes.
Eq. (\ref{eq: matrix_eq}) expresses the ME constraints for each grain in the packing.
$\vec{f}$ is a vector containing $z$ contact force magnitudes and
$A$ is a matrix which contains all of the information about the geometry, and extra rows that generate any global constraints on the force network\cite{FNE}.  The values of these global constraints are represented by $\vec b$. 
In our work, we
choose to fix all three independent elements of the stress tensor
$\sigma_{\alpha\beta}=\sum_{i=1}^{M}\sum_{j>i}^{M}r_{ij}^{\alpha}f_{ij}^{\beta}$,
so that (for 2D packings) there are three global constraints expressed
in $A$ and three non-zero constants appear in $\vec{b}$. Constructed
in this way, $A$ has $z$ columns and $2M+3$ rows. When
$A$ is rectangular with a number columns larger than the number of
rows, it will have a null space of dimension $z-2M+3$, if the
remaining $2M+3$ rows are all independent (this would not be true for crystalline packings, but will generically be true for amorphous packings). 

\section{Sampling Force Network Solutions} 
\begin{figure}
\subfigure[]{
\includegraphics[scale=0.34]{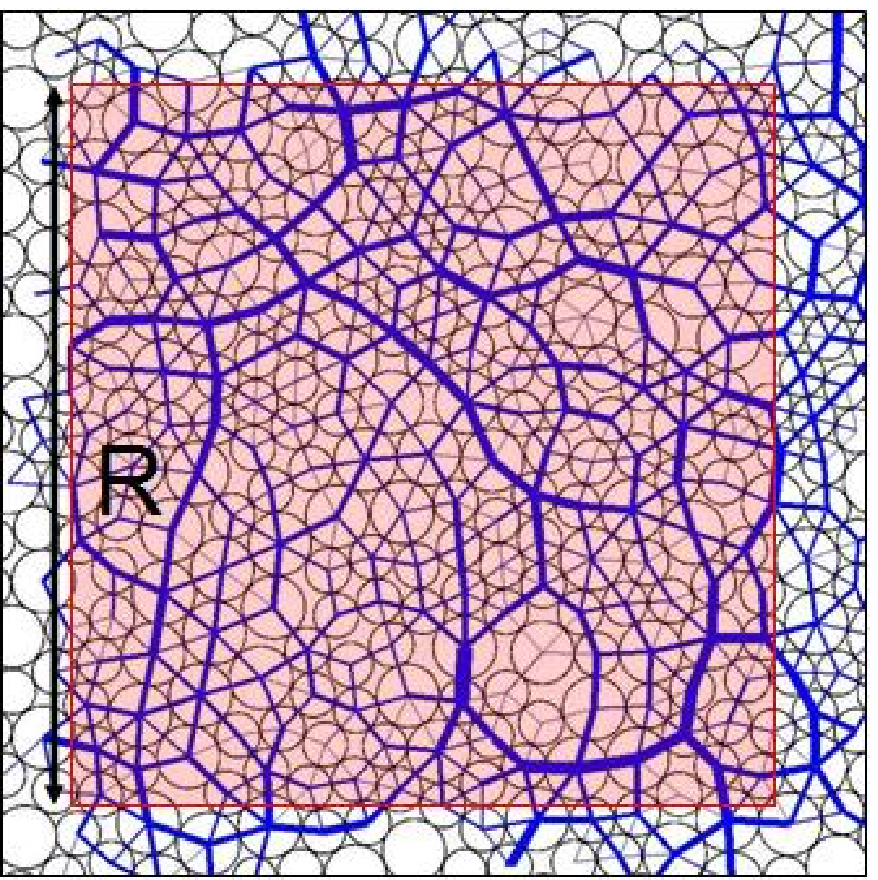}
\label{fig:PTS_example_left}
}
\subfigure[]{
\includegraphics[scale=0.35]{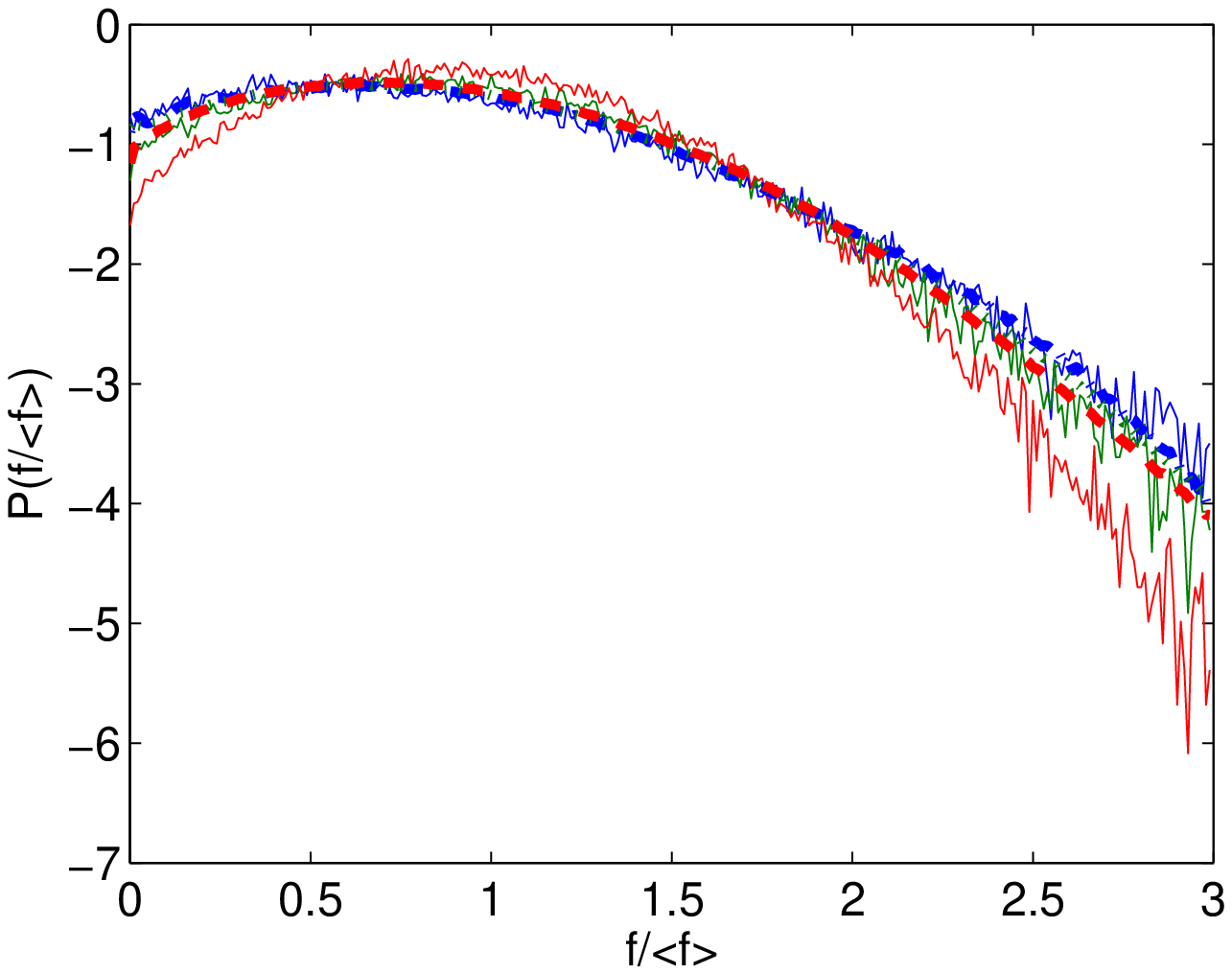}
\label{fig:PTS_example_right}
}
\caption{(a) The measurement of $C(R)$  {}``freezing'' the contact forces
on the boundary of a box of size $R$ and then sampling. (b) (solid lines) The single contact force distributions assembled from 40 different 900 grain geometries at (red) $\delta \phi = 0.1$, (green) $\delta \phi = 0.05$, and $\delta \phi = 0.01$. (dashed lines) The single contact force distributions of the FNE sampled using a single amorphous geometry at (red) $\delta \phi = 0.1$ and (blue) $\delta \phi = 0.01$.
}
\end{figure}

The null space basis vectors
$\{\vec{g}\}$ span a space of solutions for the homogeneous equation
$A\vec{f}=0$, so that these solutions are at $\sigma_{\alpha\beta}=0$.
A real packing has non-zero $\sigma_{\alpha\beta}$, so one needs
a particular solution $\vec{f}_{0}$ at the fixed $\sigma_{\alpha\beta}$.
All $\vec{f}_{0}+c\vec{g}$ with an arbitrary
coefficient c are valid force networks. We obtain $\vec{f}_{0}$ from the force law used to form the quenched geometry. Finally, while
$\vec{f}_{0}+c\vec{g}$ is a solution to eq. \ref{eq: matrix_eq},
it is not generally a solution that applies to a granular packing
because the forces are not generally positive, and the positivity constraint has to be enforced separately. 
We sample solutions by constructing a random walk in the null space of $A$. 
Assuming each valid solution
is equally likely, a new solution $\vec{f}'$ is found by choosing
a random step size $c$ from a uniform distribution, as well as a
random direction in the null space $\vec{g}$ and adding $c\vec{g}$
to a previous solution $\vec{f}$: $\vec{f}'=\vec{f}+c\vec{g}$. In
this way the sampling method is a random walk in the null space starting
at $\vec{f}_{0}$. The random walk is subject to reflecting boundary
conditions resulting from the positivity constraint on the individual
forces as well as the global constraints built into $A$, including the ones coming from the boundary that are needed to define the  PTS correlation function. 
In earlier studies of FNE\cite{WheelMove} a {}``wheel move'' has been
used to sample valid force networks.  In contrast to the wheel move, the moves
presented here are {\it intrinsically non-local}. The null space basis vectors
have of order $z$ non-zero components, so that each move $\vec{f}\rightarrow\vec{f}+c\vec{g}$
will generally change all, or most, of the contact forces simultaneously.  These non-local moves obey detailed balance, and are subject only to the constraints that are explicitly built into $A$.  We have checked that our sampling method reproduces single force distributions for real geometrically distinct packings by comparing the single force distributions of such packings to those of the FNE(see figure \ref{fig:PTS_example_right}), especially close to the unjamming transition. 

\section{Point-To-Set as a Measure of Correlations}
The PTS correlation function provides a measure of how far boundary effects penetrate into the interior of  a system\cite{PTSspin,BiroliNPhys,PTSLJ}, and is thus an indicator of a critical point.  Generically, one creates a reference configuration of random variables, such as the spins in the p-spin model, or the positions of particles in an inherent structure of the Lennard-Jones system that satisfy the constraints. A new configuration, satisfying the constraints,  is then created 
inside of some boundary $B(r)$ while outside of the boundary the
configuration is kept {}``frozen'', and an overlap function is defined to measure   
the covariance between the old and new configurations.
In the Lennard-Jones (LJ) system, for example, the PTS correlation function $q(R)$ \cite{PTSLJ} is defined as the overlap
between density functions of the reference state and the {}``pinned state,'' generated
by freezing the LJ particles outside of a boundary and allowing the
dynamics to continue inside of the boundary. 
The PTS correlation function in the LJ system exhibits a growing length scale, indicating an increasing susceptibility to boundary conditions as the liquid is cooled towards the glass transition\cite{PTSLJ}.
The jammed (mechanically stable) states of frictionless disks are inherent structures of a system with short-range repulsive interactions obtained by quenching from infinite temperature.  Keeping in mind that the only source of cohesion in these materials is imposed pressure, we would like to study the behavior of the PTS as the pressure on the packings is reduced.  
To construct the PTS correlation function for the unjamming transition in frictionless disks, we therefore study packings at a given pressure, where a region outside of a given boundary is held fixed while the interior region
is allowed to change in a way that samples new granular packings without changing the pressure. If unjamming through decompression is indeed controlled by a critical point, then the PTS correlation function should exhibit a growing length scale as the pressure is reduced to zero.

There is an inherent difficulty in constructing overlap functions for granular
packings with different geometries, and in principle different topologies. Within the FNE model, we avoid this difficulty by fixing the geometry and only allowing the contact forces to change to calculate the PTS correlation function for a given geometry. The geometry-specific PTS correlation function $\langle C(R) \rangle$ is computed by building $\vec{f}(R)$ from $\vec{f}_{0}$ with all of the contact forces outside of a radius $R$ in $\vec{f}_{0}$
left unchanged, but the contact forces inside of $R$ allowed to 
change in a way that satisfies mechanical equilibrium.   The overlap, $C(R)=\hat{f}_{0}\cdot\hat{f}(R)$, where $\hat{f}_{0}$ includes {\it only grains in the core of the region}, away from the boundary.  Since they share a single underlying
geometry, the two force networks are of the same length. The magnitudes of the core of each network are divided out so that the largest value of $C$ is $1$.  
ME equations associated with grains which share contacts that are inside of $R$ are included in the set that define $A$, however,  these particular equations cannot be set to zero.  Instead, these equations are set equal to the non-zero net forces which are the sum of all the frozen forces associated with the boundary grains.  
Two force networks are never orthogonal
to each other because of the positivity of the contact forces.  But, if we ignore the positivity constraint, the overlap of two completely uncorrelated unit vectors should simply be the projection of a randomly chosen unit vector onto one of the axes in a $z$-dimensional space.  There are $z$ components to a randomly chosen unit vector, and each component squared is roughly proportional to $1/z$, so that the overlap of two completely uncorrelated force networks would be $\propto 1/\sqrt{\frac{M_c}{M} z}$, where $M_c$ is the number of grains within the core. We expect that $C(R)$ will approach this value asymptotically at large $R$.  The overlap $C(R)$ is averaged over all $\vec{f}(R)$ sampled by the random walk in the solution space with the imposed boundary constraints to obtain $\langle C(R) \rangle$. 
To obtain the PTS correlation function as a function of pressure, $\langle C(R)\rangle$ is then averaged over many geometries at a given pressure\cite{CoreyCompressionRoutine}.  Throughout this work, we will use $\langle\rangle$ to denote an average with respect to the FNE for a fixed geometry, while $\langle\rangle_g$ will denote an average over many geometries at a given overcompression.

\section{Length Scales and Scaling} 
\begin{figure}
\onefigure[scale=0.58]{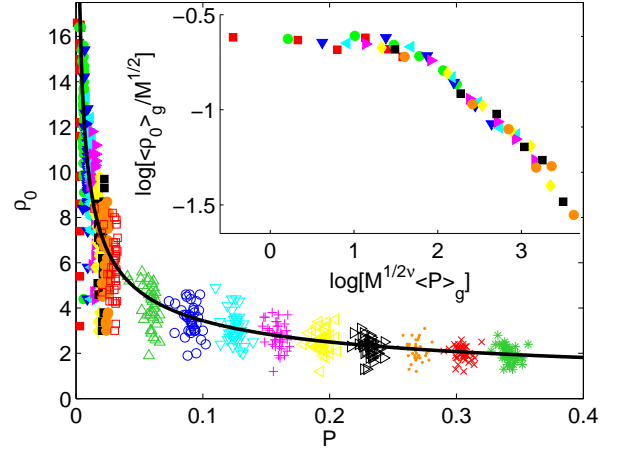}\caption{The critical subregion size $\rho_{0}$ is plotted against pressure for each of our geometries with 900 grains.  (Inset) Finite size scaling is used to collapse $\rho_0$ over the range of pressures for 150,300,450,600,750, and 900 grains, using the exponent $\nu=0.461$.  Different colors (symbols) correspond to different overcompressions, with the open symbols ranging from 0.01 to 0.1 and the filled symbols ranging from 0.001 to 0.008.  This symbol legend corresponds to the other figures of this article as well. \label{fig:FiniteSize}}
\end{figure}

\begin{figure}
\onefigure[scale=0.58]{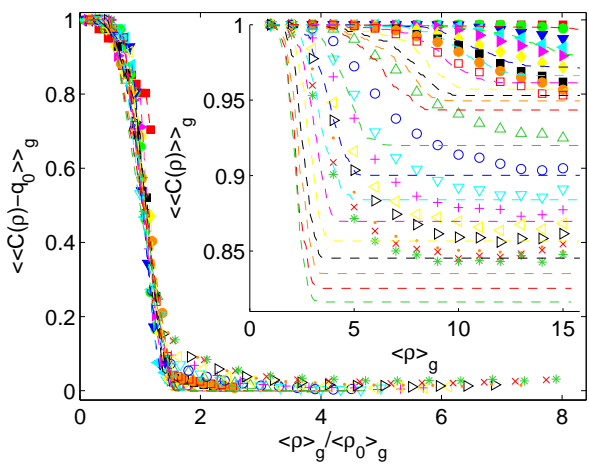}
\caption{(Inset) The measured PTS correlation function, with different symbols corresponding to different values of $<P>_g$.  The highest pressures correspond to the curves in the inset which decay to the lowest values.  The dashed lines are the predictions from eq. \ref{modelPTS}, using the measured values of $\delta n\left(\rho\right)$. In the main figure, the connected correlation function is plotted, as well as the model prediction for the connected part. \label{fig:CorrFunc}}
\end{figure}

For a series of system sizes ranging from $M=30$  to $M=900$ grains, and
overcompressions ranging from $\delta\phi=10^{-3}\ to\ 10^{-1}$,
we measured $\langle C(R) \rangle$. For each geometry,  there is a unique scale, $\rho_0 \equiv R_0/d$, below which $\langle C(\rho) \rangle=1$, precisely,  since the bounding region is small enough that the constraints on the contacts which are interior to the region cannot be satisfied except by the original network, $\hat f_0$, and there are no other solutions to the ME equations. For $R > R_0$ ($\rho > \rho_0$), the correlation function decays monotonically to its asymptotic value, $C(L/d)$.
Singular value decomposition of the matrix $A$ can be used to identify $\rho_{0}$ as the scaled bounding ball size at which the measured nullity becomes greater than $0$.   Fig. \ref{fig:FiniteSize} shows $\rho_{0}$, and its value averaged over all of the geometries at a given pressure and for a range of system sizes (measured in terms of number of grains $M$).  For the largest system size, a fit of $\langle \rho_0 \rangle_g$ versus the $\langle P \rangle_g$ yields an exponent of $\nu=$0.461$\pm$0.012. 
The inset of fig. \ref{fig:FiniteSize}, shows finite-size scaling results for $\langle \rho_0 \rangle_g$.  We obtained data collapse over five decades using the finite-size scaling form $\frac{\langle \rho _0 \rangle_g}{L} = g(L^{1/\nu}\langle P\rangle_g)$ based on $\rho_0$ diverging as $P^{-\nu}$, and taking $L \propto M^{1/2}$.  We observe the best collapse for $\nu=0.46$, with a visible failure of the collapse outside of the interval $\pm0.02$.  The figure shows only the region close to $L^{1/\nu}\langle P\rangle_g \simeq 1$,   The results shown in Fig. \ref{fig:FiniteSize} provide compelling evidence of a diverging length scale that is related to ``ordering'' in force-network space.  
The data for $\langle \langle C\left(\rho\right) \rangle \rangle_g$ for different system sizes show that for $\rho > \rho_0$, increasing the system size simply continues $\langle \langle C\left(\rho\right)\rangle \rangle_g$ to lower values with no other length scale setting in.
Fig.\ref{fig:CorrFunc} shows $\langle C\left(\rho\right) \rangle_g$, measured as a function of  $\rho=\frac{R}{d}$, the dimensionless length scale in terms of the grain diameter $d$.  As seen from the figure,  for $\rho=\frac{R}{d} << \langle  \rho_0 \rangle_g$, a pressure dependent scale,  the correlation function is identically 1. The form of $\langle \langle C(\rho)\rangle \rangle_g$ reflects the characteristics of the relevant force network solution space, and can be modeled as\cite{BiroliAdamGibbs}:
\begin{equation}
C(\rho)=p_{\vec{f}_0} (\rho) + p_{\vec{f} \ne \vec{f}_0} (\rho) q_0 ~,
\label{modelPTS}
\end{equation}
where $p_{\vec{f} \ne \vec{f}_0} (\rho)$ is the probability of finding a force network {\it different} from $\vec f_{0}$, $p_{\vec{f}_0} (\rho) = 1-p_{\vec{f} \ne \vec{f}_0} (\rho)$, and $q_0 = \langle \hat f \cdot \hat f_0 \rangle$ is the average overlap of two unequal unit vectors that are constrained by positivity, and represents the ``uncorrelated'' limit of $C(\rho)$.  A connected correlation function can, therefore, be defined as $C(\rho) - q_0 = p_{\vec{f}_0} (\rho) (1-q_0)$.    Because the number of components of the unit vectors change with pressure, $q_0$ is a function of pressure.   
The connected correlation function obtained by approximating $q_0$ by $C(\rho=L/d)$ shows  a rapid decay to zero and a collapse for different pressures after scaling $\rho$ by $\rho_0$ (Fig. \ref{fig:CorrFunc}).

To compare the predictions of the model, Eq. \ref{modelPTS}, to the measured $C(\rho)$, we need to estimate both $p_{\vec{f} \ne \vec{f}_0} (\rho)$, and $q_0$.   As argued earlier, the overlap of two uncorrelated unit force networks depends on the number of components of the force network.  Also, in the isostatic limit $z\rightarrow z_0$, $q_0\rightarrow 1$ and therefore we expect $q_0 =\sqrt{\frac{\phi_0 z_0}{\phi z}}$, where $z$ is the total number of contacts for a given pressure and using $M_c\propto\phi$.  The probability of finding a force network different from $\vec{f}_0$ depends on the volume of the solution space ($V(\rho)$) of ME force networks  for a given bounding ball size.  
The {}``force network space''  is a high dimensional space (of dimension equal to the number of contacts) and is sparsely populated by valid force network solutions. Embedded in the force network space is a solution space which is the null space of $A$ shifted to an origin defined by$\vec{f}_{0}$, and has a dimension equal to the number of  contacts in excess of the ones needed to satisfy the constraints of ME.  This space is dense with solutions, since it is spanned by the null vectors of $A$. In addition, since each null vector is an equally likely component
of a solution, the space is roughly (hyper)-spherical. Since the space is dense and spherical, the number of solutions for a given geometry and pressure goes as the volume of a hypersphere with dimension equal to the number of excess contacts, $\delta n$.  The probability of finding a force network different from $\vec{f}_0$ is, therefore, given by $p_{\vec{f} \ne \vec{f}_0} (\rho) = (V(\delta n (\rho)) -1)/V(\delta n(\rho))$, where $\delta n (\rho)$ is the average number of excess contacts for a region of size $\rho$, and we have explicitly subtracted the one coming from the single solution that is identical to $\vec{f}_0$.  The connected correlation function should, therefore, be given by: $C(\rho) - q_0 = \frac{1-q_0}{V(\delta n\rho)}$.  
In fig. \ref{fig:CorrFunc},we compare $C(\rho)$ obtained from the model (eq. \ref{modelPTS}), using measured values of $\delta n(\rho)$, and $V(\delta n (\rho)$ (discussed below) to the measured values of $C(\rho)$.  The figure also compares the connected correlation functions, which has a shape that is  very similar to that predicted by the Random First Order Transition (RFOT) theory of glasses \cite{Wolynes, BiroliAdamGibbs}.

\section{Solution Space Volume and a Mean Field Model} In order to understand the origin of the $\langle \rho_0\rangle_g$ behavior with pressure, we construct
a meanfield model based on the isocounting argument discussed earlier, 
but now including a boundary contribution \cite{TkachenkoIsocounting} to the number of constraints. If
$\delta n\left(R\right)$ is the number extra contacts within a ball
of size $R$ beyond the number of constraints (or the nullity of the
geometry matrix), then $\delta n\left(R\right)=\frac{M\delta z}{2}-B$, where $M$ is the number of grains inside of the ball, $\delta z=z-z_{iso}$
averaged over the packing, and $B$ is some boundary contribution.
Assuming homogeneity, the number of grains inside the ball is related to the overall packing
fraction: $\phi=\frac{M\pi r^{2}}{cR^{2}}\rightarrow M=\frac{cR^{2}\phi}{\pi r^{2}}=4c\phi\rho^{2}$,
where $c=\frac{1}{\left(\frac{2}{3}+\frac{1}{3}a^{2}\right)\pi}$
is the bidispersity factor, which has been kept fixed throughout this
study, with $a=1.4$. The boundary term can be estimated by first assuming that along
the border of the bounding box, there are $\sqrt{M}$ grains. Since the box has $4$ sides, and there are $2$ constraints per grain: $B=4\cdot2\cdot\sqrt{M}=16\sqrt{c\phi}\rho$, so that the meanfield expression for the nullity is:
\begin{equation}
\delta n\left(\rho\right)=2c\phi\rho^{2}\delta z-16\sqrt{c\phi}\rho
\label{eq:nullity_expression}
\end{equation}
which is nonnegative for $\rho \ge \rho_{0}=\frac{8}{\sqrt{c\phi}\delta z}$.
The critical ball size $\rho_{0}$ for which $C\left(\rho\right)$ first
becomes less than 1 is, therefore, $\frac{8}{\sqrt{c\phi}\delta z}$. A length scaling of the form $\frac{1}{\delta z}$ has been obtained earlier through an analysis of soft modes.
\cite{MatthieuLengthScale}.   
Since it is known that, on average, the deviation from isostaticity scales with the pressure $p$ as $\delta z \propto p^{1/2}$, the meanfield argument would predict that $\rho_{0}$ diverges as $p \rightarrow 0$ with an exponent of $1/2$.
This length-scale exponent has been deduced earlier from a crossover behavior of the vibrational density of states in simulations of soft-repulsive grains\cite{SilbertLiuNagel}, and the meanfield argument presented here is similar to the argument used to deduce a diverging length scale associated with isostaticity\cite{MatthieuLengthScale}.  The numerical value of the exponent deduced from the PTS correlation function is close to the meanfield value. 

\begin{figure}
\onefigure[scale=0.42]{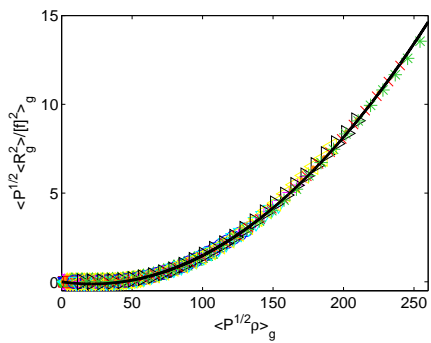}
\caption{$<<R_g^2>>_g$ collapses onto a single curve for various pressures.  Again we show data for our largest system size of 900 grains.  The black solid line is the best fit of a quadratic polynomial to the collapsed data.  \label{fig:Rg_scaling}}
\end{figure}

By its very definition, the PTS correlation function probes the size of the space of solutions that satisfy a given set of constraints. In light of the
similarity of our algorithm to a random walk in a high dimensional
space, we calculated a radius of gyration, defined as the {}``distance''
in the solution space from $\vec{f}_{0}$ to the new solution: $R_{g}^{2}=\left(\vec{f}-\vec{f}_{0}\right)^{2}$.
If the solution space is roughly spherical, this gives a measure of
the linear size of the solution space for a given quenched geometry
and bounding ball size. If the walk is in a high dimensional space,
it will spend most of its time exploring the surface of the solution
space. After some initial equilibration time required for the random
walker to reach the surface of the solution space, there is some solution
$\vec{f}_{max}$ which is {}``farthest away'' from the origin $\vec{f}_{0}$,
and if the solution space is roughly spherical, $|\vec{f}_{max}|$
is left unchanged. Taking the asymptotic value of $|\vec{f}_{max}|$, $R_{g}^{2}=\left(\vec{f}_{max}-\vec{f}_{0}\right)^{2}=\left(\sum_{i=1}^{\delta n}c_i\hat{g}_i\right)^{2}=\sum_{i=1}^{\delta n}c_i^2$, which can be approximated as $\delta n [f]^2$, where $[f]$ is the average single contact force and the square brackets are used for spatial averages over a particular geometry.
From this result we see that $R_{g}^{2}/[f]^2$ can provide a
measurement of $\delta n\left(\rho\right)$. 
The functional form of $\delta n\left(\rho\right)$
suggests a scaling form that is given by $\delta n\left(\frac{\rho}{\rho_0}\right)=\frac{1}{\delta z}G\left(\frac{\rho}{\rho_0} \right)$, with $G$ having a quadratic form.  This scaling is found
to collapse the $R_{g}^2$ data well, assuming $\delta n\left(\rho\right)=R_g^2/[f]^2$ (fig. \ref{fig:Rg_scaling}).  The master curve found from this scaling indeed is a quadratic function of $\rho$.  In (fig. \ref{fig:Rg_scaling}), The mean field equivalence between $\delta z$ and pressure is used.  The collapse illustrates the robustness of the scaling form for $\delta n$ as well as the approximation of $R_g^2$ as $\delta n[f]^2$.

The model for $C(\rho)$ (Eq. \ref{modelPTS}) involves $V\left(\delta n(\rho)\right)$, the volume of the solution space of force networks for a given $\rho$.  Assuming the space to be a hypersphere of radius $R_g (\rho)$ and dimension $\delta n (\rho)$
\begin{equation}
V\left(\delta n(\rho)\right) = e^{S\left(\delta n(\rho)\right)} =\frac{\pi^{\delta n(\rho)/2}}{\Gamma\left(\frac{\delta n(\rho)}{2}+1\right)}\left(\bar{R}_g^2(\rho)\right)^{\delta n(\rho)/2} ,
\label{eq:Entropy}
\end{equation}
where the first equality defines the configurational entropy of force networks, $S(\rho)$, which can be shown to increase linearly 
for large $\rho$ and not to have any extrema or inflection points.  The comparison in Fig. \ref{fig:CorrFunc} is based on  $V\left(\delta n(\rho)\right)$ obtained from Eq. \ref{eq:Entropy}.  The quantity $\bar{R}_g^2(\rho)\equiv\frac{R_g^2(\rho)}{R_g^2(\rho_0)}$ is is used in the definition of entropy to remove the force scale from the volume of the solution space and assure that the entropy is nonnegative.  We extract $R_g^2(\rho_0)$ directly from the numerics, and have found that it is proportional to $[f]^2$.

\section{Conclusions} We have demonstrated that a static correlation function exhibits a diverging length scale as the pressure goes to zero in frictionless granular packings.  This length scale is associated with the PTS correlation function rather than any two-point correlation function. The form of the PTS correlation function reflects the structure of the solution space of force networks, and can be obtained from the volume of this solution space.  The constraint satisfaction problem for frictionless disks is similar for frictional disks, where now there are additional constraints due to torque balance, and we expect to see a length scale in frictional materials also.   

In supercooled liquids, the PTS correlation function has been associated with the mosaic length scale \cite{Wolynes}.  In the mosaic picture, the supercooled liquid is made up of a mosaic of different metastable states.  One could speculate that in the granular systems approaching the unjamming transition, there is such a mosaic characterized by the length scale $\rho_0$ consisting of regions that are locally isostatic, but with different force network solutions.  The mosaic picture would suggest that the amorphous geometry is made up of distinct subregions, and that stress relaxation would  involve rearrangements of the mosaic tiles that overcome an entropic barrier which scale as $\rho_0$ to some power.  The time scale for stress relaxation should, therefore,  increase exponentially as the pressure decreases.  In the context of a sheared granular system, for example, where the shearing is driving the rearrangment of the distinct subregions, one would expect that the equilibration of the stress would take an exponentially long time near unjamming \cite{BehringerHowellCouette}.  We are exploring  possible ways of measuring the mosaic length scale.

\acknowledgments
This work was supported by NSF-DMR0905880, and has benefited from the facilities and staff of the Yale University Faculty of Arts and Sciences High Performance Computing Center and NSF CNS-0821132 that partially funded acquisition of the computational faciltiies.  
We acknowledge useful discussions with S. Franz, G. Biroli and Dapeng Bi, and with participants at the Les Houches Winter School on Complexity, Optimization and Systems Biology.  BC would like to acknowledge the Kavli Institute for Theoretical Physics,  where some of this work was done.

\end{document}